\documentclass[12ptl]{iopart}
\usepackage{iopams}
\usepackage{color}

\usepackage{graphicx}
\usepackage{hyperref}

\begin{document}

\title{Follow the fugitive: an application of the method of images to open
systems}
\author{G Cristadoro, G Knight and M Degli Esposti}
\address{Dipartimento di Matematica, Universit\`a di Bologna, Piazza di Porta
San Donato 5, 40126 Bologna, Italy}
 \eads{\mailto{giampaolo.cristadoro@unibo.it},
\mailto{georgiesamuel.knight@unibo.it} and  
\mailto{mirko.degliesposti@unibo.it} }


\date{\today}

\begin{abstract}
Borrowing and extending the \emph{method of images} we introduce a theoretical
framework that greatly simplifies analytical and numerical investigations of the
escape rate in open systems. As an example, we explicitly derive the exact size-
and position-dependent escape rate in a Markov case for holes of \emph{finite
size}.  Moreover, a general relation between the transfer operators of the
closed and corresponding open systems, together with the  generating function of
the probability of return to the hole is derived. This relation is then  used to
compute the small hole asymptotic behavior, in terms of readily calculable
quantities. As an example we derive  logarithmic corrections in the second order
term. Being valid for Markov systems, our framework can find application in
many areas of the physical sciences  such as information theory, network theory,  quantum Weyl law and via Ulam's method can be used as
an approximation method in general dynamical systems.
\end{abstract}

\section{Introduction}
\label{Sec:intro}

The study of the statistical properties emerging from chaotic dynamics typically
deals with equilibrium quantities and related  convergence issues, asymptotic in
time. On the other hand, there is growing interest in applications of dynamical
systems theory to problems where the dynamics is stopped or modified after a
given event occurs.  Concrete examples include leaking systems \cite{AltReview}
and metastable states \cite{Bov09}. In some cases we would like to understand
the probability of a failure event as for example in the spreading of epidemics
on networks \cite{Vesp01}. In other situations,  hitting a predetermined region
in phase space could be a desirable event, as for example if that  region is the
only accessible part from which we can gain information \cite{BuDe07}. Many of
these situations are typically modelled by \emph{open} dynamical systems which
we consider here.  A closed, discrete-time evolution  on a state space $X$ can
be opened  by defining a region $E \subset X$ in which particles can leak out,
early work on such systems can be found in \cite{PiYo79,Kad84}. If the dynamics
is chaotic enough, the measure of points remaining in the system after $n$
iterates will decrease exponentially. The escape rate, which is the rate of the
exponential decrease, is a well defined object, often admitting an
interpretation as a spectral quantity and thus invariant for a large class of
initial densities \cite{DeWrYo11,DeYo06}. On the contrary, in \cite{PaPa96}  it
was observed that varying the position of the hole has a strong effect on the
average lifetime of chaotic transients due to the complex periodic orbit
structure in chaotic maps. This dependence on the position of the hole has
generated a renovated  interest among both physicists and mathematicians; in
\cite{BuYu11} the escape rate was shown to have non-trivial dependence on the
position and non-monotonic dependence on the size of the holes, general results
on the asymptotic behavior for small holes have been studied in
\cite{KeLi09,DeWr12,Dett11,FePo11} whilst in \cite{FaEc03,AlEn10}, the addition
of noise  is investigated. Open problems and reviews of this material can be
found in \cite{DettCh,DeYo06,AltReview}. Despite the increasing interest in the
properties of the escape rate, explicit derivations remain a challenging task. 
Here we introduce a method which  greatly simplifies this task in the case
of Markov systems with a broad class of applications. The use of Markov systems
is ubiquitous in physics having applications in reaction-diffusion equations
\cite{Ko02}, chemical engineering \cite{Ta98}, meteorology \cite{Ga76}, genetics
 \cite{Tan73}, ecology \cite{Ca09}, absorbing states \cite{Nov95}, neuronal
dynamics  \cite{FrKa01}, \cite{Sch12} and quantum mechanics among others
\cite{Erm10}. In dynamical systems theory, Markov systems are widely used in
approximating more general open systems \cite{Fr01}. Via this method we
derive results  for both finite-size and  asymptotically small holes.  To this
end,  an explicit  formula that relates the transfer operators for closed and
open dynamical systems together with a return-time operator is derived.

\section{Open transfer operator: restriction to invariant subspace}
\label{Sec:operator}

While escape processes are usually interpreted in the literal sense that points
are removed from the system, we consider the following equivalent approach: a
point that hits the hole does not leave the system, but rather it is coupled to
a new point with negative mass. The two coupled particles are then evolved with
the same dynamics of the closed system and their summed contribution to the
total mass vanishes, as should be for an escaped particle.  While being a mere
reformulation of the problem, this approach makes apparent the fact that the
effect of a leak  is fully characterized by its dynamics  within the closed
system. This idea is built on the work of Lind \cite{Lind} in topological
dynamics and is reminiscent of the method of images  which can be used to solve
certain differential equations with Dirichlet boundary conditions in
electrostatics and transport problems.
Indeed, in random walk theory some absorbing boundary condition problems can be solved using the method of images \cite{FeBookI,Chetal03}. This is done via  a symmetric initial condition with  positive  and negative  mass in a corresponding system  without absorbtion. We implement such an interpretation of escape through the transfer operator of the
system. 

Consider a closed dynamics $T:X\to X$. Given a density of initial conditions
$\rho_0(x)$, the  Perron-Frobenius operator $\mathcal{L}$ associated with  $T$
evolves it to $\rho_n(x)=[\mathcal{L}^n \rho_0](x)$ after $n$ iterates. The
transfer operator of the corresponding open system is defined by
\begin{equation}
                          \mathcal{L} _{op}\rho=\mathcal{L}((1-\chi_E)\rho),
\label{Eq:Op_op}
\end{equation}
where $\chi_E$ is the indicator function of the hole $E$. The open system no
longer conserves an invariant measure and the leading eigenvalue $\lambda$
of $\mathcal{L}_{op}$ has modulus smaller than one. The corresponding
eigenfunction is the so called conditionally invariant density and
$\gamma=-\log(\lambda)$ is the escape rate  \cite{DeYo06}.

In the Ulam scheme,  the dynamics of $T$ is approximated by a Markov chain \cite{Fr99,Ba06,Boetal12}. Let $A$ be the $N \times N$ corresponding transition matrix, $N$ being the number of partition elements in the chosen level of the Ulam procedure.  Such kind of approximation has been used to study one-dimensional as well as higher dimensional dynamical systems, see for example \cite{GeDeAl12, Erm10}. Fixing $A$, that corresponds to fixing the accuracy of the Markov approximation for the closed dynamics,  we study the escape through  a Markov hole defined as an  element of the $k^{th}$
refinement of the corresponding Markov partition.  In
such situations the appropriate operator in  equation (\ref{Eq:Op_op}) is a transfer
matrix  written in the basis corresponding to \emph{all} elements of the
$k^{th}$ refinement of  the Markov partition. The size of such a matrix thus
grows exponentially with $k$. By showing that the conditionally invariant
density is piecewise-constant on a much simpler basis, we construct a transfer
matrix for the open system which grows linearly with $k$.  Note that this Markov
setting and its associated  symbolic dynamics has also a  direct
interpretation in network dynamics \cite{AfBu10} where $E$ identifies an
absorbing path on the network,  and information theory \cite{GuOd78} where for
example $E$ could correspond to a forbidden word in a communication channel.

From now on we consider $\mathcal{L}$  to be  the operator associated with  a
given level in the Ulam approximation of a closed system, 
 $I$  the corresponding Markov partition with partition parts $J_l$,
($l=1,..,N$) and $I^j$  its $j^{th}$ refinement. Fix $k\ge0$ and choose a hole
$E \in I^k$. Equivalently $E$ is  defined by a fixed word $B$ of length
$|B|=k+1$. Let $\rho_0$ be a linear combination of constant functions  on the
elements of $I^k$. From   (\ref{Eq:Op_op}) we have that
\begin{equation} 
                        \rho_1 =\mathcal{L}_{op} \rho_0= \mathcal{L} \rho_0  
-\mathcal{L} \chi_E\rho_0.
\label{Eq:density_evol}
\end{equation}
Obviously, $\mathcal{L} \rho_0$ is piecewise constant on $I^{k-1}$ and
$\mathcal{L} \chi_E \rho_0$ is constant on the set $T(E)$ and zero elsewhere.
Furthermore 
\begin{equation} 
               \rho_2=  \mathcal{L}^2 \rho_0 - \mathcal{L}^2\chi_{E}
\rho_0-\mathcal{L}\chi_E\left(\mathcal{L} \rho_0- \mathcal{L}\chi_E
\rho_0\right),
\label{Eq:density_evol_2}
\end{equation}
where $\mathcal{L}^2\chi_{E} \rho_0$ is piecewise constant on $T^2(E)$ and zero
elsewhere. More generally it is easy to see that  any function that is piecewise
constant on any $I^j$ is eventually  attracted to the space of functions that
are linear combinations of constant functions on the elements of $\{
J_1,...,J_N,T(E),...,T^{k-1}(E)\}$. Note that even if these sets are not
disjoint,  their indicator functions are linearly independent \cite{Lind}. The
space they span is left invariant by the open dynamics of (\ref{Eq:Op_op}) and
thus contains the conditionally invariant measure. We therefore restrict the
study of the open dynamics to such a space. The operator 
$\mathcal{L}_{op}$  (with $|B|>2$, but note \footnote{By defining $\phi_B(z)=0$
for $|B|=1$ in (\ref{corrpoly}), one can show that
Eqs.(\ref{Eq:zeta_finite_final},\ref{Eq:Zeta_FS}) hold for$|B|=1$ and $2$.}) 
simplifies to

\begin{equation}
\mathbf{P}=\left( \begin{array}{ccccc|cccc}\label{Eq:matrix-ssft}
& & &  &  &0 & & &\\
& & &  & & :& & &\\
& & \mathbf{A} & &   &-\alpha & & \mathbf{0}&\\
& & &  & &: & & &\\
& & &  & & 0&& &\\
\hline
& & & & & -c_1& 1&&\\
& &  \mathbf{0} & & &-c_2 & &\ddots& \\
& & & & &:&  & & 1 \\
0& .. &  1& .. & 0&-c_{|B|-2}&0 & ..&0
\end{array} \right)
\end{equation}
where  $\mathbf{A}$ is the $N\times N$ transition matrix for the closed
system
in the original partition $I$ (we assume that the closed dynamics is chaotic
enough so that $\mathbf{A}$ is aperiodic). Denoting by $ \mu_B$ and $\mu_t$ the
measure of the hole and of its parent partition element $J_t$, with respect to
the invariant measure $\mu$ of the closed system $\mathbf{A}$, we have
$\alpha=\mu_B/\mu_t$. Note that the minus signs can be interpreted as  couplings
to  points with negative mass. The $1$ in the last row is in the $r^{th}$ column
where $J_{r}$ is the partition part that the hole gets mapped onto after $k$
iterates, i.e.  $T^{k}(E)=J_{r}$.  The $c_i$ are the probabilities that points in
$E$ return to $E$ at the $i^{th}$ iteration. As we will see, an important role
is played by  the  \emph{weighted correlation polynomial}

\begin{equation}\label{corrpoly}
\phi_B(z)=1+\sum_{i=1}^{|B|-2} c_i z^{i},
\end{equation}
which is a weighted version of the one studied by Guibas and
Odlyzko\cite{GuOd81_2} and Lind \cite{Lind} in the context of topological
dynamics.
Note that the coefficients in (\ref{corrpoly}) and $\alpha$ are easily computed
in all practical situations.

\section{Deriving the dynamical zeta function}
\label{Sec:zeta}
 Not only have we reduced the growth with $|B|$ of the size of the transfer
matrix from exponential to linear, the specific form of $\mathbf{P}$ allows us
to explicitly derive the dynamical zeta function
$\zeta^{-1}_{op}(z)=\det(\mathbf{1}-z\mathbf{P})$. The logarithm of the smallest
zero of this determinant gives the escape rate. Starting from the last row of
$(\mathbf{1}-z\mathbf{P})$, multiplying by $z$ and adding to the row above and
continuing this process up to the $(N+1)^{th}$, we expand the determinant along
this row and obtain
\begin{equation}
                         \zeta^{-1}_{op}(z)= \zeta^{-1}_{cl}(z) \phi_B(z)
+\alpha z^{\left|B \right|-1} C_{t,r}(z)
\label{Eq:zeta_finite_final}
\end{equation}
where $\zeta^{-1}_{cl}(z)=\det(\mathbf{1}-z\mathbf{A})$ is the dynamical zeta
function of the closed system and
$C_{i,j}(z)=(-1)^{(i+j)}\det(\mathbf{1}-z\mathbf{A})_{ij} $ denotes the minor 
on row $i$ and column $j$.  We stress that  $\zeta^{-1}_{cl}(z)$  and $C_{i,j}(z)$  are functions of the closed system only and are thus computable independent of the choice of the hole. 
The escape rate  depends on the hole  through  its returns  (via
$\phi_{B}(z)$ and the indices $t,r$ of the parent and final partition elements) and through its size  (via $\alpha$ and $|B|$). In particular,
finite-size holes with equal measure can have different escape rates due to
different  $\phi_B(z)$ while different holes with seemingly different dynamics
can have the same escape rate if their returns  and size are equal. In the
case  of a Bernoulli shift  we have  $\zeta^{-1}_{cl}(z)=1-z$ and $C_{t,r}(z)=
\delta_{t,r}(1-z)+\mu_t z$ implying that (\ref{Eq:zeta_finite_final}) reduces
further to
\begin{equation}
               \zeta^{-1}_{op}(z)= (1-z)\tilde{\phi}_B(z) +\mu_B z^{\left| B
\right|},
\label{Eq:Zeta_FS}
\end{equation}
where $\tilde{\phi}_B(z)$ differs from $\phi_B(z)$ only in that the sum is taken
to $\left| B \right|-1$ in (\ref{corrpoly}).

\begin{figure}
\begin{center}
\includegraphics[width=12cm,height=9cm]{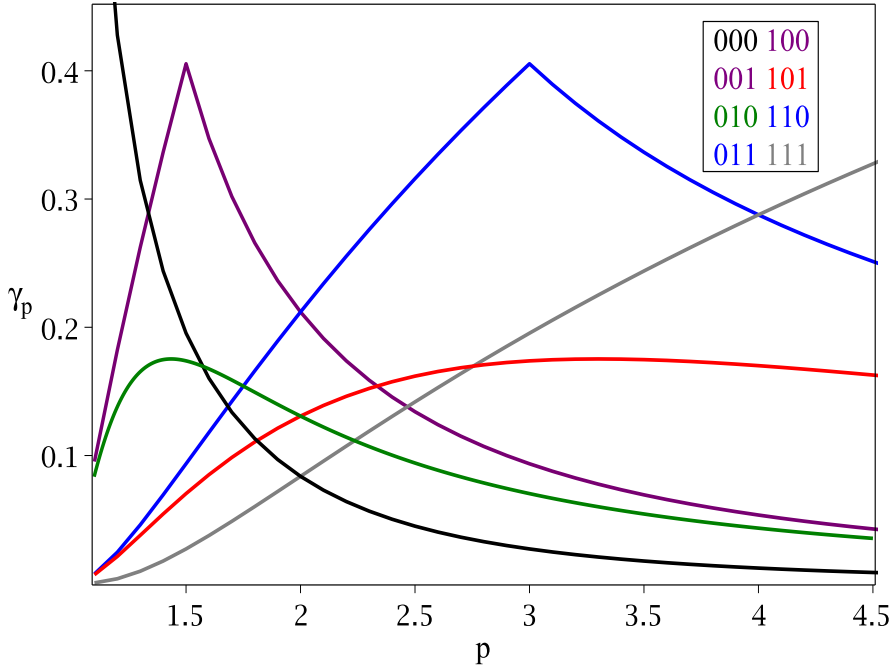}
\end{center}
\caption{\label{fig:esc_rates} (Colour online) Escape rates for cylinders of
length three in the full shift on two symbols $1,0$  with bias $p$. From $p=3$
vertically we have $000,010,001=100,101,111$ and $011=110$.}
\end{figure}

There is a vast literature on the study of dynamical zeta functions (see
\cite{ChaosBook} and references therein) but general expressions that are easy
to calculate such as  Eqs.~(\ref{Eq:zeta_finite_final},~\ref{Eq:Zeta_FS}) are
often not available \cite{Cr06}. As a particular example of the usefulness of
having an explicit zeta function, consider a Bernoulli shift on two symbols $0,1$
with transition probabilities $1/p$ and $1/q$  such that $1/p+1/q=1$. From
equation (\ref{Eq:Zeta_FS}) we can derive a relation between  the dynamical zeta
functions of the hole defined by the word $B=0^{\left| B \right|}$ and  of its
`child' hole $B_c=0^{\left| B \right|}1$ namely
\begin{equation}
               \zeta^{-1}_{0^{|B|}1}(z)=\zeta^{-1}_{0^{|B|}}(z)(1-\frac{z}{p}).
\label{Eq:left_most_hole_parent}
\end{equation}
From (\ref{Eq:left_most_hole_parent}) we can see that for some values of $p$ the
two holes have identical escape rates if the leading zero of
$\zeta^{-1}_{{0^{|B|}1}}(z)$ is equal to that of $\zeta^{-1}_{0^{|B|}}(z)$ while
 a crossover appears when the leading zero of $\zeta^{-1}_{0^{|B|}1}(z)$ comes
from the $(1-z/p)$ term, as noted in \cite{GeDeAl12}. The crossover at
$p=|B|/(|B|-1)$ is  then obtained by solving $\zeta^{-1}_{0^{|B|}}(p)=0$.
Similar results are obtained swapping  $ 0\to 1$ (i.e. $p\to q$);  see figure
\ref{fig:esc_rates}. Note that the ordering of the escape rates, for finite-size
holes, changes with $p$. At  $p=2$ (corresponding to the unbiased case) the
ordering is given by the length of the shortest periodic orbit in the holes
\cite{BuYu11}.

\section{Small hole asymptotics}
\label{Sec:asymptotics}
An additional strength of (\ref{Eq:zeta_finite_final}) is that it allows us to
study the asymptotic behavior of the escape rate as $\mu_B$ shrinks to zero
around a given point. Consider a periodic point $x\in X$ of prime period $p$
(the aperiodic case corresponds to $p\rightarrow \infty$). Pick a sequence of
shrinking holes defined by words $B_m$ chosen to be $m$ concatenated copies of a
word $w=w_1w_2...w_p$. The $w_i$ correspond to the partition sets $J_{w_i}$
visited by the prime periodic orbit of $x$.  Denote by $\mu_w$ the measure  of
the interval defined by $w$. The correlation polynomial of each $B_m$ is then
\begin{equation}\label{Eq:corr_per_final}
             \phi_{B_m}(z) = 1+\sum_{n=1}^{m-1} (c_p z^p)^{n}= \frac{1-
\mu_{B_m} z^{\left| B_m \right|}/(\mu_w \Lambda_x) }{1-z^p/ \Lambda_x}
\end{equation}
where we use $c_{kp}=c_p^k$, $\mu_{B_m}=c_p^{m-1}\mu_w$ and define
$\Lambda_x^{-1}=c_p$. For $1D$ maps $\Lambda_x$ corresponds to the stability of
the prime periodic orbit of $x$.
In order to investigate the asymptotic behavior around  $\mu_B =\epsilon \ll1$
we rewrite the smallest zero $z_0$ of (\ref{Eq:zeta_finite_final}) in terms of a
formal expansion 
\begin{equation}
                        z_0=1+\sum_{k=1}^{\infty} s_k\epsilon ^k.
\label{Eq:g_ep}
\end{equation}
Using (\ref{Eq:corr_per_final}) and (\ref{Eq:g_ep}) in
(\ref{Eq:zeta_finite_final}) along with the fact that
$\zeta^{-1}_{cl}(z)=(1-z)G(z)$ for some polynomial $G(z)$ from the
Perron-Frobenius theorem, to first order in $\epsilon$ we have,
\begin{equation}
                         \zeta^{-1}_{op}(z_0) \approx \frac{-s_1\epsilon
G(1)}{(1-\Lambda_x^{-1})}+\frac{\epsilon}{\mu_t} C(1)_{t,r}=0.
\label{Eq:zeta_zero}
\end{equation}
Which gives,
\begin{equation}
                         s_1 = \frac{C(1)_{t,r}(1-\Lambda_x^{-1})}{\mu_tG(1)}.
\label{Eq:s1}
\end{equation}
In order to make sense of (\ref{Eq:s1}) we firstly derive an important
relationship between the transfer operators of the closed and open systems and
the operator
\begin{equation}
                         T(z) =\mathbf{1}+   \sum_{n=1}^{\infty}z^n T_n
\label{Eq:T(z)}
\end{equation}
where $T_n$ is the return time operator of a set $E$ which  is defined by $T_n
\rho =\mathcal{L}^n \chi_E\rho$ similar to the operators defined in \cite{Sa02}.
From (\ref{Eq:Op_op}) we have formally     
\begin{eqnarray}\nonumber                           
\mathbf{1}-z\mathcal{L}_{op}&=&\mathbf{1}-z\mathcal{L}+z\mathcal{L}
\chi_E\nonumber\\                                                                                
        &=&(\mathbf{1}-z\mathcal{L}) \left(\mathbf{1}+   \sum_{n=1}^{\infty}z^n 
\mathcal{L}^n \chi_E\right)\nonumber\\                                                                                
        &=&(\mathbf{1}-z\mathcal{L}) T(z).
\label{Eq:Operator_relation_1}
\end{eqnarray}
We stress that this relation is generic and not restricted to the Markov setting
investigated so far. On the other hand, in the present Markov one, we can show
that $ \det(T(z))= U(z)$ is 
 the  generating function of the probability of return to the hole, while $\det(
\mathbf{1}-z\mathcal{L}_{op})=\zeta^{-1}_{op}(z)$ and
$\det(\mathbf{1}-z\mathcal{L})=\zeta^{-1}_{cl}(z)$. Finally we obtain
\begin{equation}
                         U(z)= \frac{\zeta^{-1}_{op}(z)}{ \zeta^{-1}_{cl}(z)}.
\label{Eq:Operator_relation_3}
\end{equation}
$U(z)$ is related to the  generating function of the probability of first-return
 $F(z)$ via \cite{FeBookI},
\begin{equation}
                           F(z)=\frac{U(z)-1}{U(z)}.
\label{Eq:renewal-eq}
\end{equation} 
Using (\ref{Eq:Operator_relation_3}) and (\ref{Eq:renewal-eq})  we explicitly
show that the decay rate of the first return time distribution of a set $E$ in a
closed system, given by the leading  pole of $F(z)$, is equal to the escape rate of the
system open on $E$ (given by the leading zero of $ \zeta^{-1}_{op}(z)$), a previously
heuristically derived result \cite{AltReview}. Furthermore, from Kac's lemma
$F'(1)=1/\mu_B$ and thus, using (\ref{Eq:Operator_relation_3}) in
(\ref{Eq:renewal-eq}) we have
\begin{equation}
                         \frac{1}{\mu_t}=  \frac{G(1) }{C(1)_{t,r}},
\label{Eq:t_final}
\end{equation}
which implies
\begin{equation}\label{firstOrder}
s_1=1-\Lambda_x^{-1}.
\end{equation}
Note that for $p \rightarrow \infty$ (corresponding to $x$ aperiodic) we have
$\Lambda_x^{-1}\rightarrow0$ and $s_1=1$ (equivalently  choose  $B_m$ to be
aperiodic so that $\phi_{B_m}(z)=1$ for every $m$ and the result follows).

The first order expansion (\ref{firstOrder}) is consistent with \cite{KeLi09},
wherein it is shown to be valid for more general chaotic maps. On the other
hand, using (\ref{Eq:corr_per_final}) and (\ref{Eq:t_final}) one can derive 
\emph{all} orders  recursively due to the general form of
(\ref{Eq:zeta_finite_final}). Of particular note is that for a.e. $x\in X $,
corresponding to $x$ aperiodic, we have,
\begin{equation}
                         s_2=\left| B
\right|-1-\frac{G'(1)}{G(1)}+\frac{C'(1)_{t,r}}{C(1)_{t,r}}.
\label{Eq:s2}
\end{equation}
From the Entropy (Shannon-McMillan-Breiman) theorem \cite{ShieldsBook}, we have
that for $\mu$-a.e. $x\in X$,  $\left| B \right| \approx-\ln(\epsilon)/ h$
for $\left| B \right| \rightarrow \infty$ where $h$ is the metric entropy. That
is, in the expansion of $z_0$ an $\epsilon^2 \ln(\epsilon)$ term appears for all
subshifts of finite type for  $\mu$-a.e. $x\in X $. The appearance of  an
$\epsilon^2 \ln(\epsilon)$ term is derived for the doubling map in \cite{Dett11}
along with a heuristic argument for why it should be found in a more general
setting, as we have confirmed here for the general case of a subshift of finite
type.

\section{Conclusions}
\label{Sec:Conclusions}

Having an expression for the dynamical zeta function such as 
equation (\ref{Eq:zeta_finite_final}) has allowed us to analytically derive some
properties of the escape rate for a paradigmatic class of dynamical systems.
From a mathematical viewpoint,  we believe that a formal  investigation of the
accuracy and convergence properties of the proposed framework within the Ulam
approximation scheme, could  permit the extension of some of the results to more
general nonlinear  systems. In this respect, care should be taken  as it is known that discretization can introduce `fake' eigenvalues (not related to the dynamics) even for linear hyperbolic toral automorphisms, \cite{BlKeLi02},  and similar phenomena could potentially appear in open systems. 
From a physical viewpoint,  we expect that some
of the  methods introduced could be adapted to approximate open systems in
different practical situations, such as open billiards \cite{DettCh}.
Moreover, note that the formalism is equally valid in situations where
some piecewise constant observable is introduced \cite{FePo11} or where an
infinite countable partition is used as in intermittent maps \cite{GasWan88}. 
Finally we stress that the equations (\ref{Eq:zeta_finite_final}) and (\ref{Eq:Operator_relation_3})  allow us to
extend Kac-like relations to higher moments of the probability distribution of
first-return time \cite{Hay02}: as an example, for a full shift with
$\tilde{\phi}_B(z)=0$  we find 
\begin{equation}
                          \left< \tau^2\right> = \frac{2}{\mu_B^2}-
\frac{2\left| B \right|-1}{\mu_B}.
\label{Eq:Sq_return_time}
\end{equation}
Similar expressions are obtained in the general case.  
Extending these results
will be the subject of future work. 

Furthermore, application of the ideas presented in this work, could help in studying the escape rate in more general dynamical systems.

\ack
We acknowledge partial support by the FIRB-project RBFR08UH60 (MIUR, Italy).

\section{References}


\begin{thebibliography}{10}

\bibitem{AltReview}
E~G {Altmann}, J~S~E {Portela}, and T~{T{\'e}l}.
\newblock Leaking chaotic systems.
\newblock {\em Rev. Mod. Phys.}, 85:869--918, May 2013.

\bibitem{Bov09}
A~Bovier.
\newblock Metastability.
\newblock In Roman Koteck\'y, editor, {\em Methods of Contemporary Mathematical
  Statistical Physics}, Lecture Notes in Mathematics. Springer Berlin
  Heidelberg, 2009.

\bibitem{Vesp01}
R~Pastor-Satorras and A~Vespignani.
\newblock Epidemic spreading in scale-free networks.
\newblock {\em Phys. Rev. Lett.}, 86:3200--3203, Apr 2001.

\bibitem{BuDe07}
L~A Bunimovich and C~P Dettmann.
\newblock Peeping at chaos: Nondestructive monitoring of chaotic systems by
  measuring long-time escape rates.
\newblock {\em Europhys. Lett.}, 80(4):40001, 2007.

\bibitem{PiYo79}
G~Pianigiani and J~Yorke.
\newblock Expanding maps on sets which are almost invariant: decay and chaos.
\newblock {\em Trans. Amer. Math. Soc}, 252:351--366, 1979.

\bibitem{Kad84}
L~P Kadanoff and C~Tang.
\newblock Escape from strange repellers.
\newblock {\em P. Natl. Acad. Sci. U.S.A.}, 81(4):pp. 1276--1279, 1984.

\bibitem{DeWrYo11}
M~F. Demers, P~Wright, and L-S Young.
\newblock Entropy, lyapunov exponents and escape rates in open systems.
\newblock {\em Ergod. Theor. Dyn. Sys.}, 32(04):1270--1301, 2012.

\bibitem{DeYo06}
M~F Demers and L-S Young.
\newblock Escape rates and conditionally invariant measures.
\newblock {\em Nonlinearity}, 19(2):377, 2006.

\bibitem{PaPa96}
V~Paar and N~Pavin.
\newblock Bursts in average lifetime of transients for chaotic logistic map
  with a hole.
\newblock {\em Phys. Rev. E}, 55:4112--4115, Apr 1997.

\bibitem{BuYu11}
L~A Bunimovich and A~Yurchenko.
\newblock Where to place a hole to achieve a maximal escape rate.
\newblock {\em Israel J. Math.}, 182:229--252, 2011.

\bibitem{KeLi09}
G~Keller and C~Liverani.
\newblock Rare events, escape rates and quasistationarity: Some�exact
  formulae.
\newblock {\em J. Stat. Phys.}, 135:519--534, 2009.

\bibitem{DeWr12}
M~F Demers and P~Wright.
\newblock Behaviour of the escape rate function in hyperbolic dynamical
  systems.
\newblock {\em Nonlinearity}, 25(7):2133, 2012.

\bibitem{Dett11}
C~P {Dettmann}.
\newblock Open circle maps: small hole asymptotics.
\newblock {\em Nonlinearity}, 26(1):307, 2013.

\bibitem{FePo11}
A~Ferguson and M~Pollicott.
\newblock Escape rates for gibbs measures.
\newblock {\em Ergod. Theor. Dyn. Sys.}, 32(03):961--988, 2012.

\bibitem{FaEc03}
H~Faisst and B~Eckhardt.
\newblock Lifetimes of noisy repellors.
\newblock {\em Phys. Rev. E}, 68:026215, Aug 2003.

\bibitem{AlEn10}
E~G Altmann and A~Endler.
\newblock Noise-enhanced trapping in chaotic scattering.
\newblock {\em Phys. Rev. Lett.}, 105:244102, Dec 2010.

\bibitem{DettCh}
C~P Dettmann.
\newblock Recent advances in open billiards with some open problems.
\newblock In E.~Zearoulia and J.C. Sprott, editors, {\em Frontiers in the Study
  of Chaotic Dynamical Systems with Open Problems}, World Scientific Series on
  Nonlinear Science, B, Vol. 16. World Scientific Pub. Co. Inc., 2011.

\bibitem{Ko02}
M~A Kouritzin and H~Long.
\newblock Convergence of markov chain approximations to stochastic
  reaction-diffusion equations.
\newblock {\em Ann. Appl. Probab.}, 12(3):pp. 1039--1070, 2002.

\bibitem{Ta98}
A~Tamir.
\newblock {\em Applications of Markov Chains in Chemical Engineering}.
\newblock Elsevier Science, 1998.

\bibitem{Ga76}
P~Gates and H~Tong.
\newblock On markov chain modeling to some weather data.
\newblock {\em J. Appl. Meteor.}, 15:1145--1151, Sep 1976.

\bibitem{Tan73}
W~Y Tan.
\newblock Applications of some finite markov chain theories to two locus
  selfing model with selection.
\newblock {\em Biometrics}, 29(2):pp. 331--346, 1973.

\bibitem{Ca09}
J~A Capit\'an, J~A Cuesta, and J~Bascompte.
\newblock Statistical mechanics of ecosystem assembly.
\newblock {\em Phys. Rev. Lett.}, 103:168101, Oct 2009.

\bibitem{Nov95}
M~A Novotny.
\newblock Monte carlo algorithms with absorbing markov chains: Fast local
  algorithms for slow dynamics.
\newblock {\em Phys. Rev. Lett.}, 74:1--5, Jan 1995.

\bibitem{FrKa01}
G~Froyland and K~Aihara.
\newblock Estimating statistics of neuronal dynamics via markov chains.
\newblock {\em Biol. Cybern.}, 84:31--40, 2001.

\bibitem{Sch12}
N~T Schmandt and R~F Gal\'an.
\newblock Stochastic-shielding approximation of markov chains and its
  application to efficiently simulate random ion-channel gating.
\newblock {\em Phys. Rev. Lett.}, 109:118101, Sep 2012.

\bibitem{Erm10}
L~Ermann and D~L Shepelyansky.
\newblock Ulam method and fractal weyl law for perron-frobenius operators.
\newblock {\em Eur. Phys. J. B.}, 75(3):299--304, 2010.

\bibitem{Fr01}
G~Froyland.
\newblock Extracting dynamical behavior via markov models.
\newblock In A.~I. Mees, editor, {\em Nonlinear Dynamics and Statistics}, pages
  281--321. Birkhauser Boston, 2001.

\bibitem{Lind}
D~Lind.
\newblock Perturbations of shifts of finite type.
\newblock {\em SIAM J. on Discrete Math.}, 2(3):350--365, 1989.

\bibitem{FeBookI}
W~Feller.
\newblock {\em An Introduction to Probability Theory and Its Applications},
  volume~1.
\newblock Wiley, January 1968.

\bibitem{Chetal03}
A~V Chechkin, R~Metzler, V~Y Gonchar, J~Klafter, and L~V Tanatarov.
\newblock First passage and arrival time densities for lévy flights and the
  failure of the method of images.
\newblock {\em J. Phys. A: Math. and Gen.}, 36(41):L537, 2003.

\bibitem{Fr99}
G~Froyland.
\newblock Using ulam's method to calculate entropy and other dynamical
  invariants.
\newblock {\em Nonlinearity}, 12(1):79, 1999.

\bibitem{Ba06}
W~Bahsoun.
\newblock Rigorous numerical approximation of escape rates.
\newblock {\em Nonlinearity}, 19(11):2529, 2006.

\bibitem{Boetal12}
C~{Bose}, G~{Froyland}, C~{Gonz{\'a}lez-Tokman}, and R~{Murray}.
\newblock {Ulam's method for Lasota-Yorke maps with holes}.
\newblock {\em ArXiv e-prints 1204.2329}, April 2012.

\bibitem{GeDeAl12}
O~Georgiou, C~P Dettmann, and E~G Altmann.
\newblock Faster than expected escape for a class of fully chaotic maps.
\newblock {\em Chaos}, 22(4):043115, 2012.

\bibitem{AfBu10}
V~S Afraimovich and L~A Bunimovich.
\newblock Which hole is leaking the most: a topological approach to study open
  systems.
\newblock {\em Nonlinearity}, 23(3):643, 2010.

\bibitem{GuOd78}
L~J Guibas and A~M Odlyzko.
\newblock Maximal prefix-synchronized codes.
\newblock {\em SIAM J. Appl. Math.}, 35(2):pp. 401--418, 1978.

\bibitem{GuOd81_2}
L~J Guibas and A~M Odlyzko.
\newblock Periods in strings.
\newblock {\em J. Comb. Theory A}, 30(1):19 -- 42, 1981.

\bibitem{ChaosBook}
P.~Cvitanovi\'c, R~Artuso, R~Mainieri, G~Tanner, and G~Vattay.
\newblock {\em Chaos: Classical and Quantum}.
\newblock Niels Bohr Institute, Copenhagen, 2010.
\newblock {\tt \href{http://ChaosBook.org}{ChaosBook.org}}.

\bibitem{Cr06}
G~Cristadoro.
\newblock Fractal diffusion coefficient from dynamical zeta functions.
\newblock {\em J. Phys. A: Math. and Gen.}, 39(10):L151, 2006.

\bibitem{Sa02}
O~Sarig.
\newblock Subexponential decay of correlations.
\newblock {\em Invent. Math.}, 150:629--653, 2002.

\bibitem{ShieldsBook}
P~C Shields.
\newblock {\em Graduate Studies in Mathematics}, volume~13.
\newblock Amer Mathematical Society, 1996.

\bibitem{BlKeLi02}
M~Blank, G~Keller, and C~Liverani.
\newblock {R}uelle-{P}erron-{F}robenius spectrum for {A}nosov maps.
\newblock {\em Nonlinearity}, 15(6):1905, 2002.

\bibitem{GasWan88}
P~Gaspard and X-J Wang.
\newblock Sporadicity: Between periodic and chaotic dynamical behaviors.
\newblock {\em P. Natl. Acad. Sci. U.S.A.}, 85(13):4591--4595, 1988.

\bibitem{Hay02}
N~Hadyn, J~Luevano, G~Mantica, and S~Vaienti.
\newblock Multifractal properties of return time statistics.
\newblock {\em Phys. Rev. Lett.}, 88:224502, May 2002.

\end{thebibliography}

\end{document}